# Strong Confinement of optical fields using localised surface phonon polaritons in cubic Boron Nitride

Ioannis Chatzakis[1], Athith Krishna[2], James Culbertson[3], Nicholas Sharac[1], Alexander J. Giles[3], Michael G. Spencer[2], and Joshua D. Caldwell[3,4]

[1]*ASEE/NRC Postdoctoral fellow (residing at NRL, Washington, DC)*
[2]*School of Electrical and Computer Engineering, Cornell University, Ithaca, NY 14853*
[3]*U.S. Naval Research Laboratory, Washington, DC 20375*
[4]*Dept. of Mechanical Engineering, Vanderbilt University, Nashville, TN 37212*
*Corresponding author: ioannis.chatzakis.ctr.gr@nrl.navy.mil, josh.caldwell@vanderbilt.edu*

**Phonon polaritons (PhPs) are long-lived electromagnetic modes that originate from the coupling of infrared photons with the bound ionic lattice of a polar crystal. Cubic-Boron nitride (cBN) is such a polar, semiconductor material, which due to the light atomic masses can support high frequency optical phonons. Here, we report on random arrays of cBN nanostructures fabricated via an unpatterned reactive ion etching process. FTIR reflection spectra suggest the presence of localized surface PhPs within the Reststrahlen band, with quality factors in excess of 38 observed. These can provide the basis of next generation infrared optical components like antennas for communication, improved chemical spectroscopies, and enhanced emitters, sources and detectors.**

The identification of surface phonon polaritons (SPhPs) as a low loss alternative to surface plasmons has led to a surge in recent research identifying polar dielectric materials capable of supporting these deeply sub-diffractional modes. These long-lived electromagnetic modes originate from the coupling of infrared photons with the bound ionic lattice vibrations of a polar crystal. These polaritons have been demonstrated to provide exceptional increases in the quality factor ($Q$-factor) for localized polariton modes in comparison to their plasmonic (SPP) counterparts. Due to their phononic origin, the scattering lifetimes of SPhPs are orders of magnitudes longer than those of SPPs [1]. These advantages have been experimentally verified and exploited within several polar dielectric materials, notably silicon carbide [1-4], boron nitride nanotubes [5], hexagonal boron nitride (hBN) [6], III –nitride [7], $SiO_2$ [8- 10], and III-V semiconductors [11,12].

Two key applications of interest for these materials are surface enhanced infrared absorption (SEIRA) [12-17] for molecular sensing and solid-state, narrow-band thermal emitters for infrared optical sources [18]. Highly selective, strong thermal emission with narrow bandwidth, controlled polarization [19,20] and spatial coherence [18] in the infrared are required for realizing sources/ detectors with higher efficiency and for targeted applications [2].

While the materials discussed above span a spectral range including the mid/long-wave and far-infrared, a spectral gap exists between the active regions of hBN and SiC where no high quality SPhP materials have been reported (Fig. 1). This spectral region, between ~ 972 and 1370 $cm^{-1}$, is critical for molecular sensing as it correlates with the energy of vibrations for phosphate, phosphene, sulfate, and nitro functional groups, which are found in many chemical and biological warfare agents and explosive compounds. Therefore, identifying an appropriate SPhP material that could be used for SEIRA-based detection schemes is of significant interest.

Cubic boron nitride (cBN) is one material that can potentially meet these needs. The so-called Reststrahlen band, bound by the transverse (TO) and longitudinal optic (LO) phonons, is found between ~1050 and 1300 $cm^{-1}$, thus, SPhPs can be supported over a spectral range covering much of the regime desired.

A dramatic modification of the optical response of materials (e.g. molecules) can occur when they are placed in the vicinity of nanostructures that induce strong, highly localized electromagnetic fields [2]. While SPhP nanostructures have been fabricated via e-beam lithography, a more commercially viable method is needed. Here we demonstrate lithography-free patterning of randomly distributed cBN nanostructure arrays, realizing localized resonances with *Q*-factors on par with the best plasmonic systems. Our method, which relies only on reactive ion etching (RIE) is highly scalable, accessible, and relatively inexpensive, in comparison to e-beam lithography. We show that the aperiodic nanostructures support highly absorptive, localized SPhP resonances suitable for SEIRA and tailored thermal emitters. Despite the random orientation and large variations in nanoparticle sizes via this lithography-free method, the SPhP resonances still exhibit linewidths of 32 cm$^{-1}$, offering a Q-factor of 38, higher than the best pure plasmonic resonators [21]. Here Q is defined as the ratio of the resonant frequency to the full-width half-maximum of the resonance intensity spectrum ($\omega/\Delta\omega$). High Q indicates a low rate of the energy loss relative to the energy stored in the resonator, and thus has been used as a figure of merit. However, the Q-factor does not consider the modal volume nor incorporate the confinement of electromagnetic energy, which is the key point for nanophotonic devices. In order to include the physical concept of the confinement and the respective modal volumes of the SPhP modes, the Purcell factor ($F_p \propto \lambda^3_{res}/V_{eff}$) needs to be considered. This describes the radiation-cavity coupling as a function of the optical properties of the cavity. Based on the geometrical characteristics of the particles inferred by Atomic Force Microscopy (AFM) and Scanning Electron Microscopy (SEM) images, and the wavelength of the resonance $\lambda_{res}$, we estimate modal volumes ~7.5 times smaller than the $\lambda^3_{res}$ and predict very large $F_p$ in the order of 103. This implies a strong enhancement of the emission rate of nearby emitters in this spectral range. Fig. 2 shows the morphology and reflectivity of the etched cBN sample. The etched sample surface shows significant variation between regions, with particles shapes and sizes formed as shown in the SEM images (Fig. 2a-c). The highest Q for the localized SPhP resonances are observed in regions where the nanoparticles are predominantly oval-shaped (Fig. 2a), as demonstrated in the corresponding reflectivity spectra shown in Fig. 2d. Particles with trigonal or tetrahedral shapes were also observed, but typically exhibited a broader linewidth, with values approaching 50 cm$^{-1}$. The dramatic variation in particle morphology may be due to inhomogeneity of the etching process. Possible explanations for etching variability may be that the surface of the cBN crystals was initially not uniformly smooth, or that it was covered by an uneven density of adsorbents. Similar tetrahedral shape particles have also been obtained by Guo *et al.* (2014) on cBN surface having introduced in the mixture CaF2 and MgF2 [22]. They found that the relative growth rate of tetrahedral particles depends on the etchant and the exposed crystal facet. In particular, in the (100) facet etched by LiF, the tetrahedral islands were formed much faster than on the (111) facet, and the etch rate in the (111) direction by CaF2 and MgF2 was faster than in the (100).

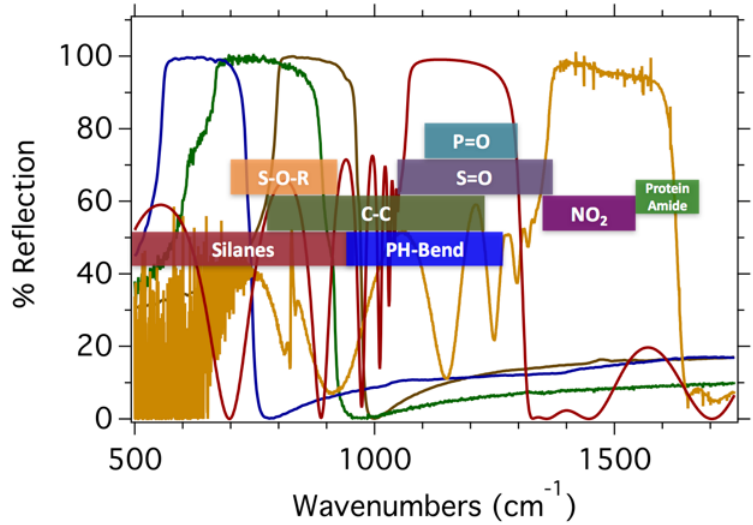

**Fig. 1.** The reflectivity spectrum in the Reststrahlen band for GaN (blue), AlN (green), SiC (brown), cBN (red), and hBN (yellow). The horizontal colored bars represent the molecular vibrations of a plethora of compounds that can easily be identified using sensors based on the corresponding material. Most of the compounds have their vibrational frequencies in the reststrahlen band of the cBN.

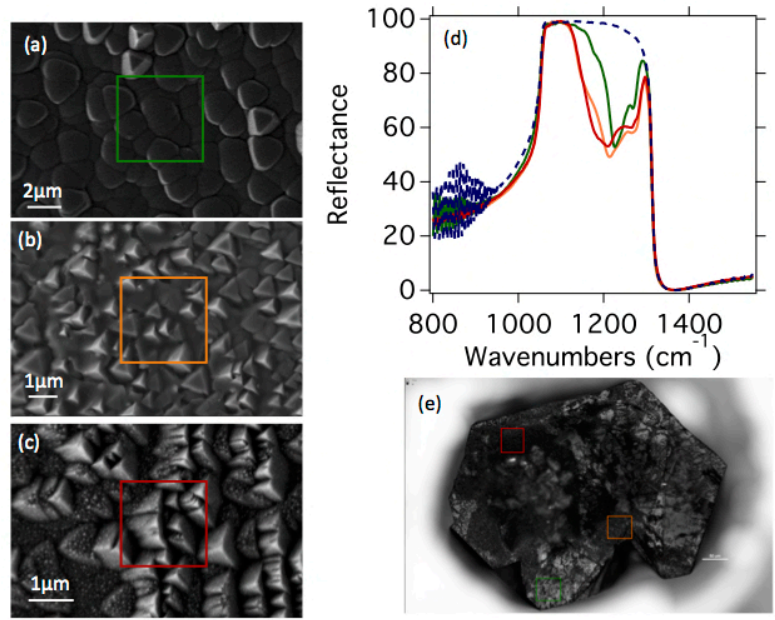

**Fig. 2.** Scanning Electron Microscopy (SEM) images (a), (b), (c) and (e) for different etched areas on the 200 μm size cBN grain shown in panel (e). The corresponding FTIR reflectance spectra of each area in the cBN sample are shown in panel (d) with the same color of panels (a) – (c.) The blue dashed line is the reflectance of unpatterned cBN.

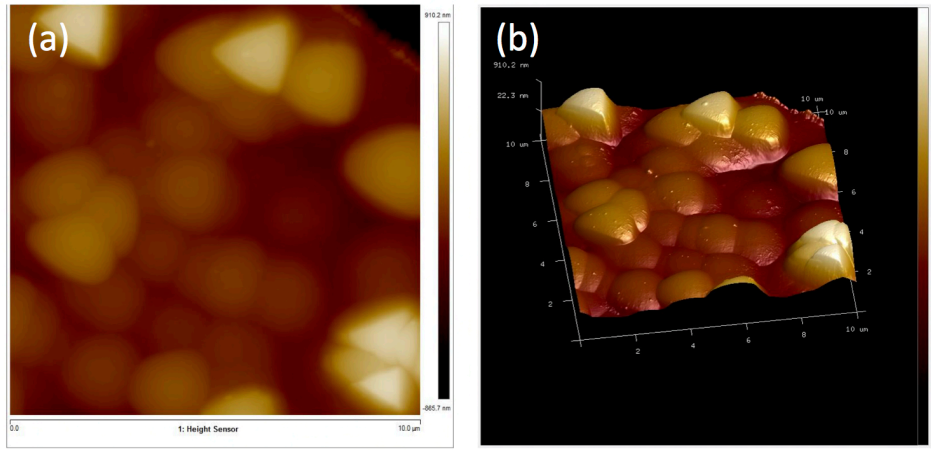

**Fig. 3.** Representative AFM images of a selective probed area. (a) Height map where the differences in the shape and dimensions of the particles can be identified. The scale bar represents the height of the particles. (b) Isometric view of data from (a).

Here we used CF$_4$ and SF$_6$ for the plasma etching. The AFM images in Fig. 3a, b show the morphology of the area that produces the highest Q resonances (see green line spectra in Fig. 2d). In order to verify the origin of the observed resonances as SPhP in nature, we employed electromagnetic simulations using the CST Studio Suite. Prior to this, the complex dielectric function of cBN was required. To derive the complex dielectric permittivity, we modeled the reflection spectra shown in Fig. 5a using a single –Lorentzian form (Eq. (1)) of the dielectric tensor components [23]:

$$\varepsilon(\omega) = \varepsilon_\infty \frac{\omega_{LO}^2 - \omega^2 - i\gamma_{LO}\omega}{\omega_{TO}^2 - \omega^2 - i\gamma_{TO}\omega} \qquad (1)$$

where $\varepsilon_\infty$ is the high frequency dielectric permittivity, $\omega_{LO}/\omega_{TO}$, are the frequencies of the LO/TO optical phonon modes and $\gamma_{LO}/\gamma_{TO}$ are the corresponding damping rates. The initial guess for these parameters was taken from the literature [24-26] and final values extracted using a least-square fitting routine within the Wvase software provided by J.A. Woolam Inc. The real and the imaginary part of the dielectric function extracted from the fit are plotted in Fig. 4b. From the fit of the experimental data we determined frequencies of 1053 cm$^{-1}$, and 1306 cm$^{-1}$ for the TO and LO phonon modes, respectively, and 3.62 cm$^{-1}$ and 4.76 cm$^{-1}$ for the associated damping constants. In addition, we carried out Raman scattering measurements where the corresponding values of the TO, and LO phonon modes are 1056 cm$^{-1}$, and 1305.5 cm$^{-1}$, and the associated linewidths are 5.65 cm$^{-1}$ and 6.31 cm$^{-1}$, corresponding to ~0.94 and 0.84 ps lifetimes, respectively, similar to those measured in naturally abundant hBN [27]. This demonstrates that with improved growth and fabrication there is even further promise in cBN-based SPhP approaches. The calculated dielectric function used in the simulations is presented in Fig. 4b. The experimental Reststrahlen band collected from the unetched cBN crystals was reproduced by the simulations.

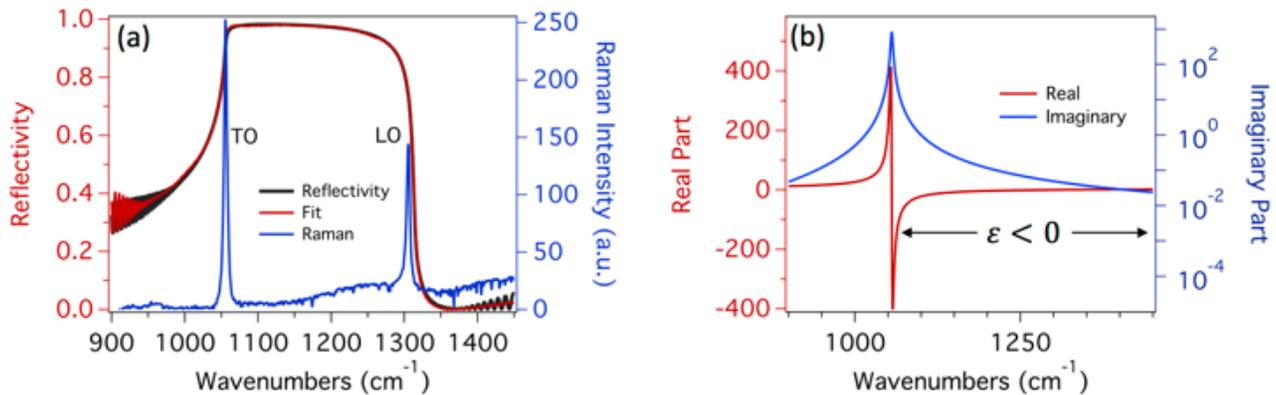

**Fig. 4.** Panel (a) presents the FTIR reflectivity spectrum (black line), in addition to the Raman spectrum (blue line). The positions of the TO and LO peaks in the Raman spectrum define the boundaries of the Reststrahlen band where the reflectivity is maximized. Panel (b) shows the real (red line) and the imaginary parts (blue line) of the dielectric function extracted from the fit of the experimental data of panel (a).

In an effort to model the nanostructures, we focused on the highest $Q$ regions with the nanoparticles estimated as ellipsoidal cylinders. In particular, in our simulations we parameterized the long and short axis of the elliptical shape pillars using 18 different permutations in the range of 1.8 – 2.5 µm for the long axis, and two values (0.9 µm and 1.05 µm) for the short axis based on SEM imaging. All the above spectra were averaged and plotted in Fig. 5. Qualitatively speaking, a possible explanation for the multiple resonances observed in the spectral range between 1200 cm$^{-1}$ to 1300 cm$^{-1}$ in the simulated data, could be the existence of monopole and transverse dipolar modes, similar to those observed by Caldwell *et. al* (2013) [4]. In addition, the effect of the inter-particle distance on the resonances is also important. Chen *et al*. (2014) [27] demonstrated that as the distance between particles decreases the energies of the monopole and transverse dipole resonances blue and red shift, respectively. However, within a region with a wide variety of resonance frequencies and linewidths, the resulting spectra will be composed of the superposition of all resonances present. Thus, such a result cannot be approximated quantitatively without including a much broader range of permutations. In our sample this distance indeed tends to zero. Such strong interparticle interactions will induce further linewidth broadening, exacerbating this effect. In our calculations the distance between the particles was fixed at the value of 2.5 µm, so the monopole and dipole resonances were spectrally resolved. Due to the spectral location of the Reststrahlen band of cBN, sensors can be developed based on this material for detecting chemical compounds with their vibrational modes in the same frequency domain, such as phosphates and nitride compounds (see Fig. 1). Also, great enhancement in the detection sensitivity can be obtained by the particle-induced resonances patterned as it has been reported above. While in many applications the random nature of the nanoparticle size and shape would limit its usefulness, for chemical sensing this is not necessarily the case. In SEIRA, enhancement will occur when spectral overlap between the molecular vibrational resonances and the localized SPhP occurs. This enhancement is thus maximized with a narrow SPhP resonance linewidth. However, this precludes the opportunity to have broader spectral coverage to couple to a number of different molecular vibrational modes. In contrast, a very low $Q$ resonator will provide broader spectral coverage, but with weak enhancement.

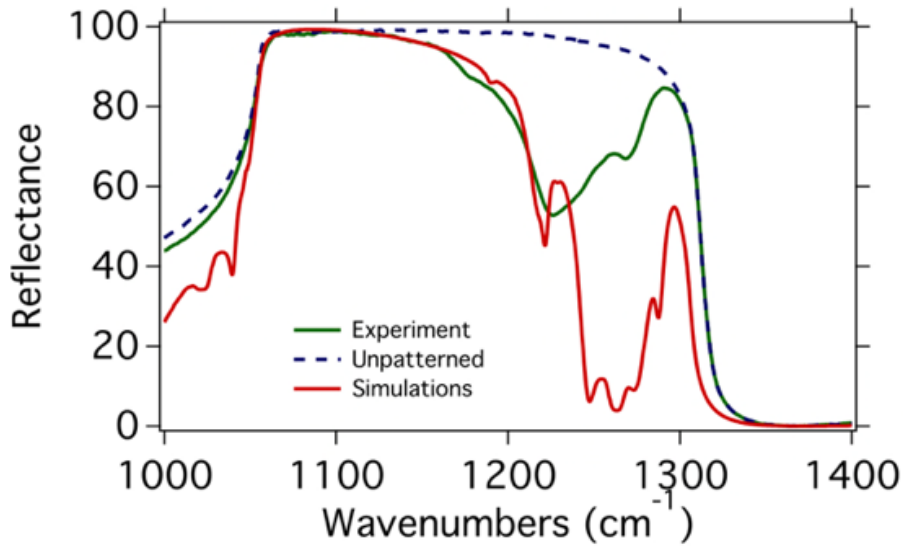

**Fig. 5.** FTIR reflectance spectrum of the randomly distributed particles etched on cBN (green solid line) compared to the simulated spectrum (red solid line). The blue dashed line is the reflectivity spectrum of the unetched cBN material.

A nice compromise for this application can be found in an array of relatively high $Q$ resonators with varying sizes, causing the full resonant spectra to be a broad superposition of a number of relatively narrow resonances. Under such conditions, strong enhancements can occur with the resonant particles that are spectrally coincident with a given molecular vibrational mode, but the large number of distinct particle shapes and corresponding resonance frequencies can offer the broader spectral coverage of a low-$Q$ cavity if the entire array is probed. The ability to realize such spectral response via a scalable approach such as the one we present here, only enhances the opportunities for these methods. In addition to the benefits for chemical sensing, there also exist opportunities within thermal emission, metameterials, and nonlinear optics in the IR and THz spectral range, presumably provided that improvements in the particle size dispersion were achieved.

In summary, we demonstrate the realization of cBN nanostructures by using a reactive ion etching method to create aperiodically distributed particles that produce resonance $Q$- factors in excess of what is typically observed with monodisperse, top-down fabricated SPP-based structures. These resonances are derived from the localized SPhPs that are supported within the nanostructures inside the Reststrahlen band. We show that light can be confined in volumes significantly smaller than the incident wavelength, with *Q-factor*s as high as 38. These results suggest a lithography-free approach to fabricating nanostructures in cBN that can effectively confine light in the subdiffraction limit in the IR regime by localized SPhP resonances. The advantage of the long-lived SPhPs, in contrast with SPP lifetimes, enables nanophotonic devices with enhanced IR absorption, extremely sensitive detection capabilities for molecular spectroscopy, and negative refractive index materials.

**Methods:** For the sample preparation commercially available High Pressure High Temperature grown cBN powder was used. A single cBN particle ~200 μm was immobilized on a Silicon substrate via graphitization of a photoresist. Oxygen plasma treatment was then used to clean the top surface of the cBN particle prior to etching. The sample was then exposed to a $CF_4$ plasma for 13 min, and then a $SF_6$ plasma for an additional 13 min. Another oxygen plasma treatment was carried out to remove remaining residual organics. The sample with the <111> plane exposed to the plasma will develop triangular, pyramidal and oval shape nanoparticles. To characterize the resonances a thermoscientific Nicolet FTIR microscope was used with a 15x objective 0.58 NA. The SEM and AFM images were taken by a Zeiss LEO and Bruker AFM microscope, respectively. The Raman spectra were collected by a home-built system with a 50x Mitutoyo objective lens at 488 nm wavelength excitation, and 2 mW power with a spot diameter 0.5 μm. To simulate the resonances in the reflectance spectrum we numerically solved the Maxwell equations by performing 3-D electromagnetic calculation using CST Studio Suite software. All of the parameters for the model have been extracted from the SEM/AFM images and the dielectric function of the cBN has been determined by fitting the experimental data of the reflectivity from the unpatterned cBN sample. For the simulations, P- and S-polarization were employed such that both the transverse and the longitudinal components of the electromagnetic fields relative to the pillars were included. All the resultant spectra from both polarizations were averaged and shown in Fig. 4.

**Funding sources and acknowledgments.** Funding was provided by the Office of Naval Research and distributed by the Nanoscience Institute at the Naval Research Laboratory. The author I. Chatzakis acknowledges support from the American Society of Engineering (ASEE) NRL Postdoctoral Fellowship Programs, and N. S. acknowledges support from the National Research Council (NRC). Funding for A. K. and M. G. S. was provided by the National Science Foundation (NSF)

**Funding.** National Science Foundation (NSF) (ECCS-1542081); NSF MRSEC program DMR-1719857